\documentclass[aps,pra,preprint,superscriptaddress]{revtex4}
\usepackage{graphicx}
\usepackage{amsmath}
\usepackage{amssymb}
\usepackage{bm} 
\usepackage{epsf}
\usepackage[english]{babel}
\usepackage{dcolumn}
\usepackage{color}

\begin{document}
\title{X-ray induced electron and ion fragmentation dynamics in IBr}

\author{Phay J. Ho}
\author{Dipanwita Ray}
\altaffiliation{Present address: AEye Corporation, Dublin, California 94568}
\author{Stefan Lehmann}
\altaffiliation{Present address: Advanced Research Center for Nanolithography ARCNL, Science Park 106, NL-1098XG Amsterdam, Netherlands}
\author{Adam E. A. Fouda}
\author{Robert W. Dunford}
\author{Elliot P. Kanter}
\author{Gilles Doumy}
\affiliation{Chemical Sciences and Engineering Division, Argonne National Laboratory, Lemont, Illinois 60439}
\author{Linda Young}
\affiliation{Chemical Sciences and Engineering Division, Argonne National Laboratory, Lemont, Illinois 60439}
\affiliation{The James Franck Institute and Department of Physics, The University of Chicago, Chicago, Illinois 60637}
\author{Donald A. Walko}
\affiliation{Advanced Photon Source, Argonne National Laboratory, Lemont, Illinois 60439}
\author{Xuechen Zheng}
\author{Lan Cheng}
\affiliation{Department of Chemistry, Johns Hopkins University, Baltimore, Maryland 21218, USA}
\author{Stephen H. Southworth}
\affiliation{Chemical Sciences and Engineering Division, Argonne National Laboratory, Lemont, Illinois 60439}

\date{\today}

\begin{abstract}
Characterization of the inner-shell decay processes in molecules containing heavy elements is key to understanding x-ray damage of molecules and materials and for medical applications with Auger-electron-emitting radionuclides. The 1s hole states of heavy atoms can be produced by absorption of tunable x-rays and the resulting vacancy decays characterized by recording emitted photons, electrons, and ions. The 1s hole states in heavy elements have large x-ray fluorescence yields that transfer the hole to intermediate electron shells that then decay by sequential Auger-electron transitions that increase the ion's charge state until the final state is reached. In molecules the charge is spread across the atomic sites, resulting in dissociation to energetic atomic ions. We have used x-ray/ion coincidence spectroscopy to measure charge states and energies of I$^{q+}$ and Br$^{q'+}$ atomic ions following 1s ionization at the I and Br \textit{K}-edges of IBr. We present the charge states and kinetic energies of the two correlated fragment ions associated with core-excited states produced during the various steps of the cascades.  To understand the dynamics leading to the ion data, we develop a computational model that combines Monte-Carlo/Molecular Dynamics simulations with a classical over-the-barrier model to track inner-shell cascades and redistribution of electrons in valence orbitals and nuclear motion of fragments. 
\end{abstract}


\maketitle

\section{INTRODUCTION\label{intro}}

Radiation damage limits the use of x-ray scattering and diffraction for structural determinations of macromolecular crystals and biological matter \cite{Stern09:ACD,Hemonnot17:ACSN}. A primary damage mechanism is x-ray absorption by inner-shell electrons causing the emission of photoelectrons and the creation of inner-shell vacancies that decay by radiative and radiationless transitions. For materials containing atoms of intermediate or higher atomic numbers Z, x-ray emission is an important decay mode of \textit{K}-shell vacancies \cite{Krause79A:jpcrd}. X-ray emission transfers the \textit{K}-shell vacancy to intermediate shells that then decay by Auger-electron emission \cite{Dunford12:pra,Southworth19:pra}. Higher-Z atoms contain several electron shells, resulting in vacancy decays by alternative pathways and wide ranges of ion charge states, e.g., Kr$^{q+}$ (q = 2-8) and Xe$^{q+}$ (q = 4-11) \cite{Southworth19:pra}. In molecules, delocalized electronic states, charge transfer, and non-local decays (interatomic Coulombic decay and electron-transfer-mediated decay~\cite{Jahnke2020:cr}) spread charge across the molecule and it dissociates into energetic atomic ions \cite{Dunford12:pra,Southworth19:pra}. Radiation damage from core-hole decays can be used to advantage in medical therapies by using radiopharmaceuticals to place radionuclides near cancerous tissues so that emitted Auger electrons destroy the DNA and membranes of tumorous cells \cite{Howell20:ijrb,Ku19:rc}. Fundamental studies of x-ray and inner-shell processes are needed for understanding radiation damage mechanisms and for applications to medical therapies.

To reduce the complexity of core-hole decays, coincidence measurements among ejected photons, electrons, and ions can separate and identify competing pathways \cite{Arion15:jesrp,Bomme13:rsi,Kukk17:as,Guillemin18:pra}. We previously used synchrotron x rays and x-ray/ion coincidence spectroscopy to study effects of pre-edge \textit{K}-shell resonances in Kr, Xe, and XeF$_2$ \cite{Southworth19:pra}. Ion fragmentation of XeF$_2$ following Xe 1s $\rightarrow$ 7$\sigma_u$ resonant excitation and Xe 1s ionization was also studied~\cite{Southworth19:pra}. Here we apply the same instrumentation and methods to measure ion charge states and dissociation energies following ionization of IBr at the Br and I \textit{K}-edges. To better understand the decay dynamics, we develop a computational model that combines Monte-Carlo/Molecular Dynamics simulations \cite{Ho14:prl,Ho15:pra} with a classical over-the-barrier model to track inner-shell cascades and redistribution of electrons in valence orbitals and nuclear motion of the fragments. The calculations provide a time-dependent model of the electron orbital hole populations, charge states, and nuclear separations for different decay pathways. Our method allows us to distinguish electron redistribution models. A recent application of the calculational methods to x-ray spectroscopy of a solvated transition metal complex is given in Ref. \cite{Chelsea21:jcp}.

Multiphoton ionization of molecules using intense, ultrashort x-ray free-electron lasers has revealed novel ionization mechanisms~\cite{Rudenko17:nat,Li21:prl} and molecular structures have been recorded by Coulomb explosion imaging~\cite{Boll22:np}. The measurements and calculations reported in the present paper are for $K$-shell photoionization by single x rays followed by radiative and radiationless transitions determined by internal electron-photon and electron-electron interactions~\cite{Crasemann85}. However, ionization by intense x rays and by vacancy cascades following absorption of hard x rays both lead to high charge states, charge transfer, and energetic fragmentation.

Section~\ref{expt} of this paper describes the experimental instrumentation and methods, Section~\ref{theory} describes the theoretical and calculational methods, and Section~\ref{res} discusses the measured and calculated results.  Conclusions and suggestions for future research are given in Section~\ref{con}.

\section{EXPERIMENTAL METHODS\label{expt}}

\begin{figure}
\includegraphics[trim=3cm 3cm 3cm 3cm,clip,width=8.6cm]{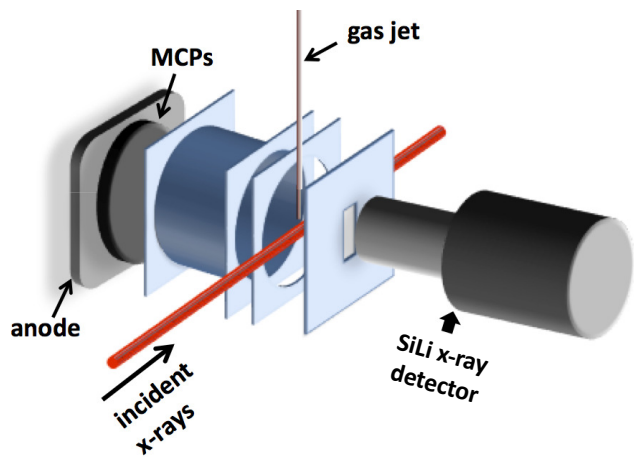}
\caption{(Color online) Schematic diagram of the x-ray/ion coincidence instrument. A SiLi x-ray detector and ion time-of-flight (iTOF) spectrometer are positioned on opposite sides of the x-ray beam and are parallel to the x-ray linear polarization. The incident x rays pass through an effusive gas jet and create an interaction region between the x-ray and ion spectrometers.  The data acquisition system records iTOF spectra in coincidence with fluorescent x rays recorded by the SiLi detector.}
\label{schematic}
\end{figure}

The experiments were conducted on beamline 7-ID at the Advanced Photon Source \cite{Walko16:cp} using the x-ray/ion coincidence spectrometer described in Ref. \cite{Southworth19:pra} and illustrated in Fig. \ref{schematic}. Horizontal and vertical slits upstream of the instrument defined the x-ray beam dimensions to $<$1$\times$1 mm$^2$. The incident x rays passed through an effusive jet of IBr molecules and created an interaction region between the SiLi x-ray spectrometer and ion time-of-flight (iTOF) spectrometer. The vertical and horizontal dimensions of the interaction region were defined by the x-ray beam dimensions and a 2.5 mm slit along the beam direction passed x rays emitted from the interaction region to the SiLi detector. The iTOF is used with static potentials, beginning with push-pull elements that direct the ions into a drift tube. The ion detector consists of a Z-stack of microchannel plates (MCPs) and a two-dimensional, position-sensitive delay-line anode. The incident x-ray energy was tuned to either the Br or I $K$-edge and x-ray absorption produced $K$-shell vacancies. The fluorescence yields of \textit{K}-shell vacancies in Br and I are 0.618 and 0.884, respectively \cite{Krause79A:jpcrd}. X-ray emission transferred the \textit{K}-shell vacancy to the $L_{2,3}$, $M_{2,3}$ or $N_{2,3}$ subshells which then decayed by series of Auger transitions and charge transfers.  The vacancy decays produced ranges of final charge states and resulted in dissociation to energetic atomic ions \cite{Southworth19:pra,Dunford12:pra}. Detection of an emitted x ray by the SiLi detector triggered the data acquisition circuit that recorded the emitted x-ray energy, the Br and I ion times-of-flight and their anode positions. Events in which initial \textit{K}-shell vacancy states decayed by Auger transitions were not recorded.

\subsection{Calculated Ionization Energies}

To calculate the Br 1s and I 1s ionization energies of IBr, we began with the delta-coupled-cluster singles and doubles method augmented with a noniterative inclusion of triple excitations [$\Delta$ CCSD(T)]~\cite{Zheng22:pccp,Raghavachari89:cpl}. Scalar-relativistic effects were treated using the spin-free exact two-component theory in its one-electron variant (SFX2C-1e)~\cite{Dyall01:jcp,Liu09:jcp}. Corrections for nuclear size, scalar two-electron picture-change effects, the spin-orbit corrections, the Breit-term contributions, and the QED effects~\cite{Koziol18:jcp} are listed in Table I. These high-order corrections were included in the same way as our previous calculations on Kr, Xe, and XeF$_2$~\cite{Southworth19:pra}. The calculated Br 1s ionization energy is 1.5 eV lower than the photoelectron measurement on HBr~\cite{Boudjemia20:pccp} and the calculated I 1s energy is 1.4 eV higher than the measurement on CH$_3$I~\cite{Boudjemia19:pccp}. The 1s ionization energies of IBr were not directly measured in the present work but are expected to be within 1-2 eV of the calculated energies listed in Table~\ref{table1}.

\begin{table}
\caption[]{\label{table1} Calculated Br 1s and I 1s ionization energies of IBr (in eV).}
\begin{ruledtabular}
\begin{tabular}{lcc}
  & Br & I  \\ \hline
SFX2C-1e-Delta-CCSD(T)   & 13502.9 & 33267.1 \\
$\Delta$[nuclear size]  &    -0.2 &    -2.7  \\
$\Delta$[scalar 2e-pc] &     8.2 &    30.8 \\
$\Delta$[2e-SO(DC)]    &   -0.3    &  -2.6  \\
$\Delta$[Breit]    &   -19.6   &   -75.9   \\
$\Delta$[QED]   &   -10.3   &   -40.2   \\
\hline
 Total                  &  13480.6 & 33176.6 \\
 Experiment             & 13482.1(3)$^a$ & 33175.2$^b$   \\
\end{tabular}
\end{ruledtabular}
\begin{flushleft}
$^a$HBr~\cite{Boudjemia20:pccp}.
$^b$CH$_3$I~\cite{Boudjemia19:pccp}.
\end{flushleft}
\end{table}

\subsection{X-ray Energies}

For measurements near the $\sim$13.4 keV Br \textit{K}-edge, the undulator's 3rd harmonic was used with a diamond 111 double-crystal monochromator. For measurements near the $\sim$33.2 keV I \textit{K}-edge, the undulator's 5th harmonic was used with the diamond 333 reflection \cite{Walko16:cp}. The estimated bandwidths at the Br and I edges were both $\sim$1 eV. The flux near 13.4 keV was $\sim$2 $\times$ 10$^{12}$ x-rays/sec and the flux near 33.2 keV was $\sim$2 $\times$ 10$^{10}$ x-rays/sec. To calibrate the energies of the incident x-rays precisely, ion yields were recorded across the Br and I \textit{K}-edges. The 4p$^5$ configuration of atomic Br results in strong 1s $\rightarrow$ 4p$\sigma^*$ pre-edge resonances in the Br \textit{K}-shell x-ray absorption spectra of Br$_2$ \cite{Filipponi98:jcp}, CF$_3$Br \cite{Fouda20:jpb}, and HBr \cite{Boudjemia20:pccp}. The 4p$\sigma^*$ resonance in the ion yield scan of IBr was recorded at $\sim$13474 eV, which is 2 eV lower than the resonance energy in CF$_3$Br \cite{Fouda20:jpb} and $\sim$1 eV higher than the resonance energy in Br$_2$ \cite{Filipponi98:jcp}. For the present measurements we tuned the incident energy to 13486 eV in order to be above the Br 1s ionization energy of IBr. 

After calibration of the x-ray energies at the Xe \textit{K}-edge \cite{Southworth19:pra}, the inflection point of the I \textit{K}-edge scan of IBr was located at $\sim$33165 eV. The I absorption edge is broadened by a $\sim$10 eV lifetime width \cite{Krause79B:jpcrd,Boudjemia19:pccp} and by pre-edge resonances. Analysis of the x-ray absorption coefficient of atomic and molecular iodine vapor shows a pre-edge resonance at 33171 eV and an ionization edge at 33176 eV \cite{Gomilsek08:jpb}. This energy matches measurements using x-ray photoelectron spectroscopy of the I 1s ionization energies of CH$_3$I and CF$_3$I that are 33175.2 eV and 33175.9 eV, respectively \cite{Boudjemia19:pccp}. The present measurements on IBr were made at $\sim$33194 eV in order to be above the I 1s ionization energy.

\subsection{X-ray/Ion Coincidence}

\begin{figure}
\includegraphics[trim=3cm 0cm 4cm 0cm,clip,width=8.6cm]{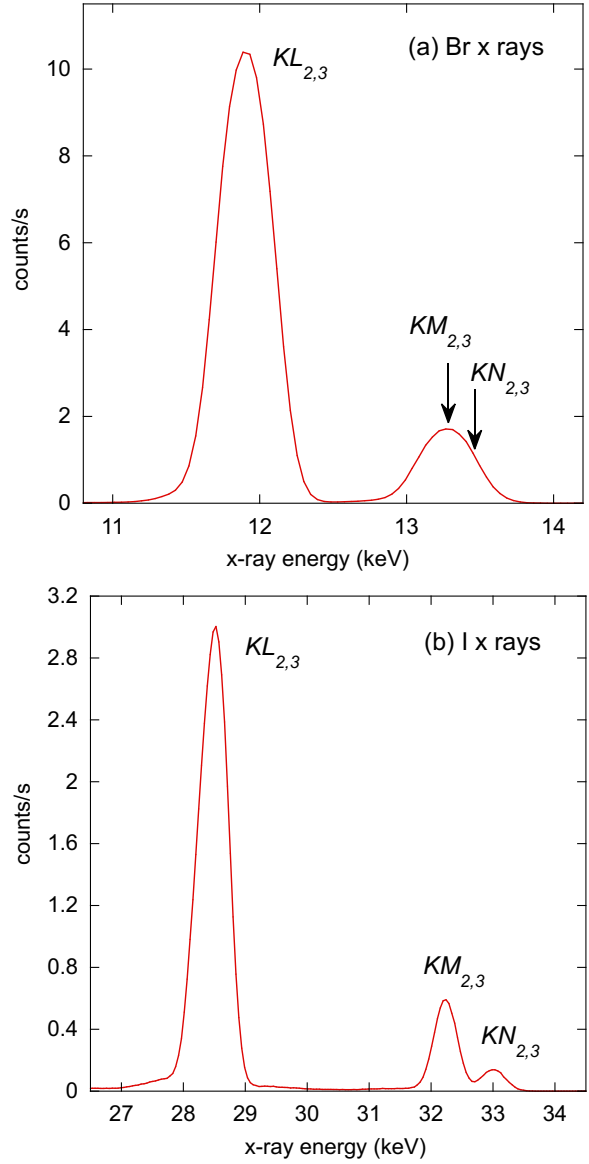}
\caption{(Color online) X-ray emission spectra of IBr measured with a SiLi detector using the x-ray/ion coincidence setup illustrated in Fig.~\ref{schematic}. (a) Br x rays generated by absorption of 13486 eV x rays. (b) I x rays generated by absorption of 33194 eV x rays.}
\label{XES}
\end{figure}

X-ray emission spectra from Br and I were recorded by the SiLi detector and are plotted in Fig.~\ref{XES}. The data analysis selects ion TOF spectra that are in coincidence with particular radiative transitions by placing filters on the x-ray emission energy range. The results discussed in this report are limited to iTOF measurements in coincidence with $KL_{2,3}$ ($K\alpha_{1,2}$) x rays in which initial 1s vacancies are transferred to the 2p shell of the same Br or I atomic site. The 2p holes then decay by series of Auger electron emission steps with charge redistribution and participation of delocalized electrons in the latter stages of the decay cascade. The result is to spread charge between the two sites leading to dissociation to energetic atomic ions.

\begin{figure}
\includegraphics[trim=0cm 3cm 1cm 3cm,clip,width=8.6cm]{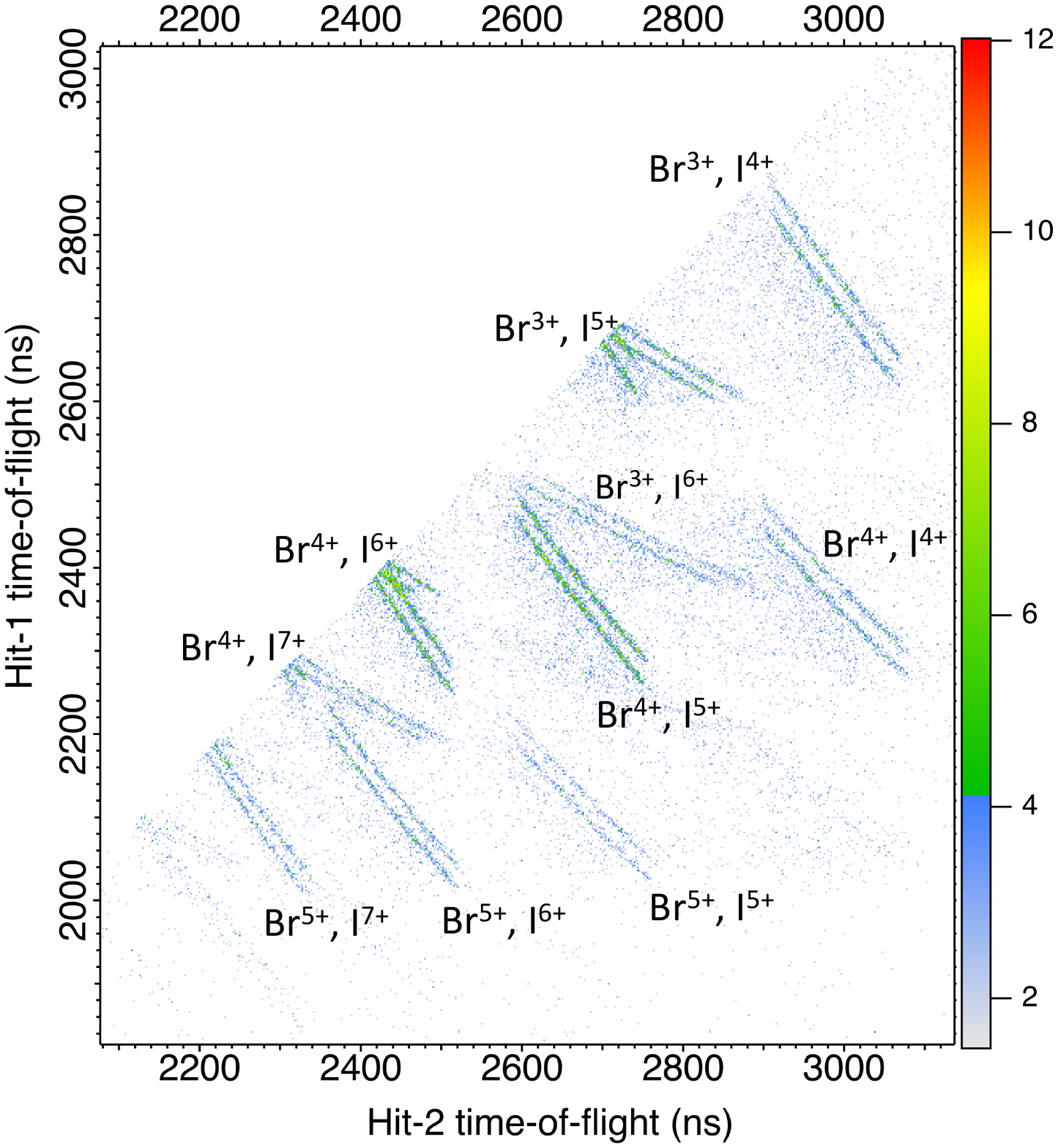}
\caption{(Color online) Time-of-flight scatter plot of Hit-1 vs Hit-2 for IBr ion fragmentation events following photoionization with 33194 eV x rays. The ion data are in coincidence with an I \textit{K}$\alpha_{1,2}$ photon. For each pair of charge states, the double lines result from Br isotopes 79 and 81 in coincidence with the single I isotope 127.}
\label{H1H2}
\end{figure}

\begin{figure}
\includegraphics[trim=0cm 0cm 0cm 2cm,clip,width=8.6cm]{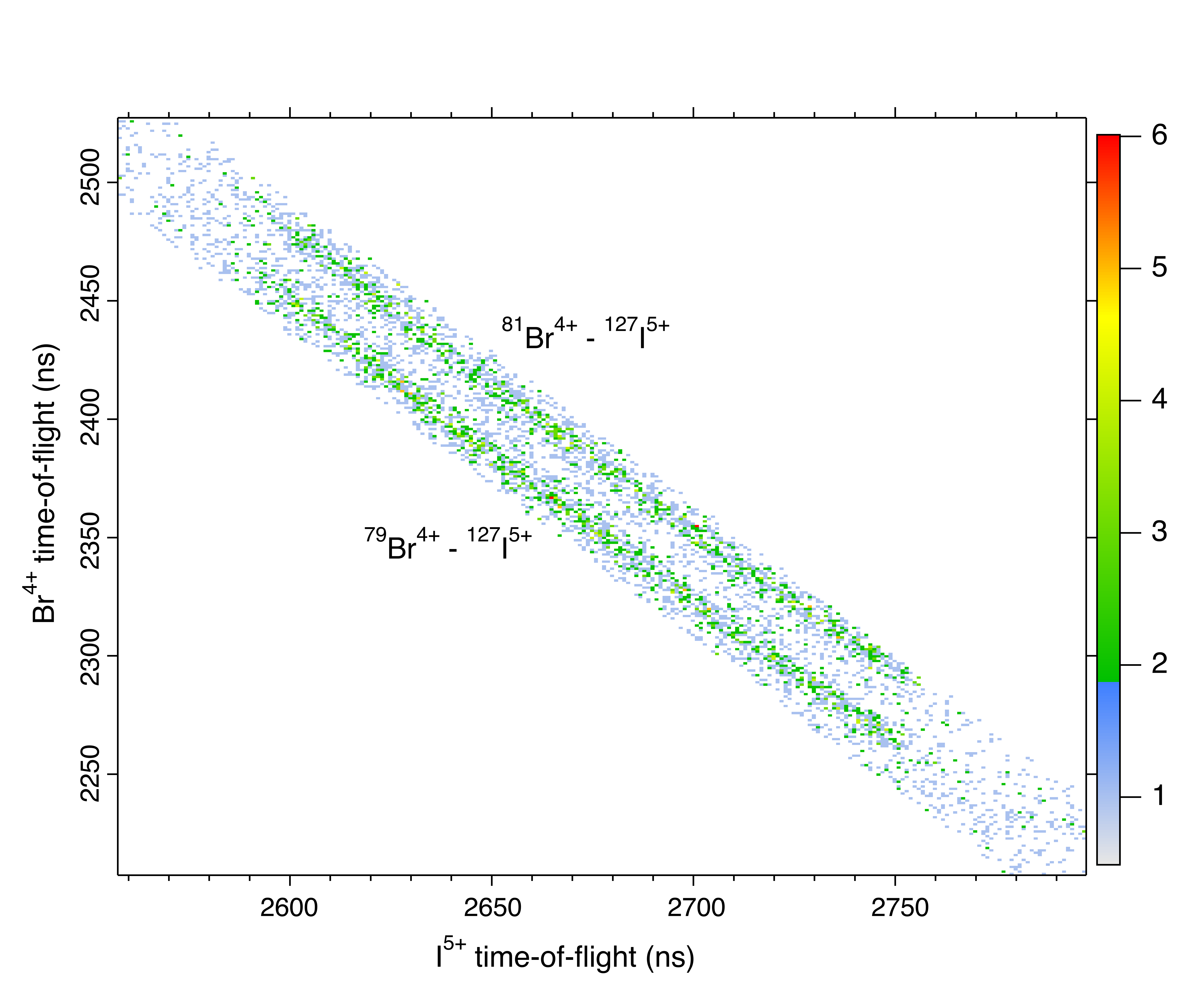}
\caption{(Color online) Time-of-flight scatter plot of Hit-1 vs Hit-2 for ion fragmentation of IBr into Br$^{4+}$ and I$^{5+}$ following photoionization with 33194 eV x rays. The data are in coincidence with I \textit{K}$\alpha_{1,2}$ photons. The double lines result from Br isotopes 79 and 81 in coincidence with the single I isotope 127. The spreads in the TOFs reflect the dissociation energies of the ions.}
\label{Br4I5}
\end{figure}

Since each valid event produces a correlated pair of Br and I atomic ions that dissociate energetically due to the Coulombic potential energy, the TOFs of the pair are recorded in the scatter plot of Hit-1 vs. Hit-2 in Fig.~\ref{H1H2}. The charge-state pairs appear as parallel lines due to the Br isotopes 79 and 81 in coincidence with the single I isotope 127. Several of the charge-state pairs are labeled in Fig.~\ref{H1H2}. The data in Fig.~\ref{H1H2} were recorded at the I $K$-edge and accumulated over 7 hours. A similar data set at the Br $K$-edge was collected over 6 hours. The projections of the lines on the Hit-1 and Hit-2 axes correspond to the TOF ranges resulting from dissociation of molecules randomly oriented with respect to the ion spectrometer axis. The components of the ion momenta parallel to the axis result in longer or shorter TOFs due to trajectories away from or toward the detector, respectively. As discussed in Ref.~\cite{Southworth19:pra} for ion fragmentation of XeF$_2$, the dissociation energy can be determined from the time required for an ion ejected away from the detector to turn around in the electric field of the iTOF spectrometer \cite{Wiley55:rsi}. We thus have a type of photoion-photoion-coincidence (PIPICO) spectrometer \cite{Arion15:jesrp} triggered by coincidence with x-ray fluorescence. By simulating the geometry and potentials of the field plates, drift tube, and detector, the TOFs of the ions were calculated, including the turn-around times that account for the ion dissociation energies. Events from each charge-state pair were isolated in the data analysis and the spreads in the TOFs used to determine dissociation energies. An example is shown in Fig.~\ref{Br4I5} for breakup into Br$^{4+}$ and I$^{5+}$. From the projections onto the iTOF axes, the ion energies are estimated to be 74-eV $^{79}$Br$^{4+}$ with 44-eV $^{127}$I$^{5+}$ and 73-eV $^{81}$Br$^{4+}$ with 44-eV $^{127}$I$^{5+}$. 

\begin{table}
\caption[]{\label{table2} Kinetic energies (in eV) of Br and I charge states following photoionization of IBr at 13486 eV and in coincidence with Br $K\alpha$ x-ray emission. The estimated uncertainties of the energies are 5\%. The kinetic energy release (KER) is the sum of the two ion energies. For comparison, Coulomb energies are calculated at the neutral ground state internuclear distance of 0.2469 nm~\cite{Huber}. The p-Br/p-I column is the ratio of the ion momenta. The rows are ordered by Br isotope and charge.}
\begin{ruledtabular}
\begin{tabular}{cccccccc}
Br isotope & Br charge & Br energy & $^{127}$I charge & I energy & KER & Coulomb energy & p-Br/p-I \\
\hline
79 & 2$^+$ & 14.4 & 2$^+$ & 9.5 & 23.8 & 23.3 & 0.97 \\
79 & 2$^+$ & 21.1 & 3$^+$ & 13.4 & 34.5 & 35.0 & 0.99 \\
79 & 3$^+$ & 20.9 & 2$^+$ & 14.0 & 34.9 & 35.0 & 0.96 \\
79 & 3$^+$ & 32.3 & 3$^+$ & 19.4 & 51.7 & 52.5 & 1.02 \\
79 & 3$^+$ & 46.1 & 4$^+$ & 28.1 & 74.2 & 70.0 & 1.01 \\
79 & 4$^+$ & 29.2 & 2$^+$ & 18.8 & 48.0 & 46.7 & 0.98 \\
79 & 4$^+$ & 43.5 & 3$^+$ & 26.2 & 69.7 & 70.0 & 1.01 \\
79 & 4$^+$ & 53.7 & 4$^+$ & 33.8 & 87.5 & 93.3 & 0.99 \\
81 & 2$^+$ & 14.0 & 2$^+$ & 9.4 & 23.4 & 23.3 & 0.98 \\
81 & 2$^+$ & 21.4 & 3$^+$ & 14.1 & 35.5 & 35.0 & 0.98 \\
81 & 3$^+$ & 21.1 & 2$^+$ & 14.3 & 35.4 & 35.0 & 0.97 \\
81 & 3$^+$ & 31.2 & 3$^+$ & 19.9 & 51.2 & 52.5 & 1.00 \\
81 & 3$^+$ & 44.1 & 4$^+$ & 29.2 & 73.3 & 70.0 & 0.98 \\
81 & 4$^+$ & 27.5 & 2$^+$ & 19.3 & 46.8 & 46.7 & 0.95 \\
81 & 4$^+$ & 43.5 & 3$^+$ & 27.8 & 71.4 & 70.0 & 1.00 \\
81 & 4$^+$ & 55.5 & 4$^+$ & 34.2 & 89.6 & 93.3 & 1.02 \\
\end{tabular}
\end{ruledtabular}
\end{table}

From the iTOF data recorded above the Br $K$-edge at 13486 eV and in coincidence with Br $K\alpha$ x rays, eight ion pairs were analyzed for both Br isotopes $^{79}$Br and $^{81}$Br ions in coincidence with $^{127}$I ions. The ion charge states, dissociation energies, kinetic energy releases (KERs), and the ratios of the momenta of the Br and I ions are tabulated in Table~\ref{table2}. The ratios of the momenta provide a check on the energy determinations since the ratios are expected to equal 1. For comparison with the measured KERs, Coulomb energies are calculated at the neutral ground state internuclear distance of 0.2469 nm~\cite{Huber}. Based on reproducibility of the determinations of the ion turn-around times, we estimate the uncertainties in the dissociation energies at the Br $K$-edge to be $\sim$5\%. Similarly, twelve ion pairs were analyzed in data recorded above the I $K$-edge at 33194 eV in coincidence with I $K\alpha$ x rays and those results are listed in Table~\ref{table3}. The data rates were smaller at the I $K$-edge, and the estimated uncertainties in the ion energies are larger, $\sim$10\%.

\begin{table}
\caption[]{\label{table3} Kinetic energies (in eV) of Br and I charge states following photoionization of IBr at 33194 eV and in coincidence with I $K\alpha$ x-ray emission. The estimated uncertainties of the energies are 10\%. The kinetic energy release (KER) is the sum of the two ion energies. For comparison, Coulomb energies are calculated at the neutral ground state internuclear distance of 0.2469 nm~\cite{Huber}. The p-Br/p-I column is the ratio of the ion momenta. The rows are ordered by Br isotope and charge.}
\begin{ruledtabular}
\begin{tabular}{cccccccc}
Br isotope & Br charge & Br energy & $^{127}$I charge & I energy & KER & Coulomb energy & p-Br/p-I \\
\hline
79 & 2$^+$ & 29.3 & 4$^+$ & 17.2 & 46.4 & 46.7 & 1.03 \\
79 & 3$^+$ & 33.5 & 3$^+$ & 22.2 & 55.7 & 52.5 & 0.97 \\
79 & 3$^+$ & 45.7 & 4$^+$ & 28.4 & 74.0 & 70.0 & 1.00 \\
79 & 3$^+$ & 55.6 & 5$^+$ & 33.2 & 88.8 & 87.5 & 1.02 \\
79 & 3$^+$ & 68.1 & 6$^+$ & 43.9 & 112 & 105 & 0.98 \\
79 & 4$^+$ & 63.6 & 4$^+$ & 34.1 & 97.7 & 93.3 & 1.08 \\
79 & 4$^+$ & 73.6 & 5$^+$ & 44.0 & 118 & 117 & 1.02 \\
79 & 4$^+$ & 90.0 & 6$^+$ & 55.7 & 146 & 140 & 1.00\\
79 & 4$^+$ & 106 & 7$^+$ & 57.8 & 164 & 163 & 1.07 \\
79 & 5$^+$ & 85.9 & 5$^+$ & 54.6 & 140 & 146 & 0.99 \\
79 & 5$^+$ & 108 & 6$^+$ & 75.1 & 183 & 175 & 0.95 \\
79 & 5$^+$ & 122 & 7$^+$ & 81.9 & 204 & 204 & 0.96 \\
81 & 2$^+$ & 28.8 & 4$^+$ & 16.9 & 45.7 & 46.7 & 1.04 \\
81 & 3$^+$ & 32.6 & 3$^+$ & 21.1 & 53.7 & 52.5 & 0.99 \\
81 & 3$^+$ & 44.0 & 4$^+$ & 28.4 & 72.3 & 70.0 & 0.99 \\
81 & 3$^+$ & 54.7 & 5$^+$ & 32.1 & 86.8 & 87.5 & 1.04 \\
81 & 3$^+$ & 59.9 & 6$^+$ & 38.9 & 98.8 & 105 & 0.99\\
81 & 4$^+$ & 57.4 & 4$^+$ & 38.9 & 96.2 & 93.3 & 0.97 \\
81 & 4$^+$ & 73.0 & 5$^+$ & 44.0 & 117 & 117 & 1.03 \\
81 & 4$^+$ & 83.1 & 6$^+$ & 57.2 & 140 & 140 & 0.96 \\
81 & 4$^+$ & 107 & 7$^+$ & 61.7 & 169 & 163 & 1.05 \\
81 & 5$^+$ & 95.0 & 5$^+$ & 60.1 & 155 & 146 & 1.00 \\
81 & 5$^+$ & 110 & 6$^+$ & 63.9 & 174 & 175 & 1.05 \\
81 & 5$^+$ & 121 & 7$^+$ & 86.3 & 207 & 204 & 0.94 \\
\end{tabular}
\end{ruledtabular}
\end{table}

\section{THEORY\label{theory}}

The x-ray induced electron and molecular dynamics in a heavy-element-containing molecule is a complex process. After a localized x-ray excitation, the molecule undergoes multistep vacancy cascades, including fluorescence and Auger decays, that transfer the vacancies in the inner-shells to valence shells.  This process involves creating transient intermediates with different electronic configurations and  charge states.  Once the vacancies reach the valence shells, electron transfer can take place to create charged atomic sites, leading to molecular fragmentation.  This deexcitation process spans multiple timescales from sub-femtoseconds to picoseconds.

To capture the dynamics leading to the ion dissociation, we developed a theoretical model that combines Monte-Carlo/Molecular Dynamics simulations \cite{Ho2017:jpb} with a classical over-the-barrier (COB) model \cite{Schnorr2014:prl} to track inner-shell cascades and redistribution of electrons in valence orbitals and nuclear motion of fragments.  The COB model has previously been applied to molecular fragmentation dynamics in intense EUV \cite{Schnorr2014:prl} and x-ray fields \cite{Erk2014:science,Rudenko2017:nature}.

We track the cascade process with a Monte Carlo method, which determines the quantum electron transition probabilities of all participating electronic configurations (ECs), including ground state, core-excited states and valence-excited states of all charge states. The electronic structure theory is based on the relativistic Hartree-Fock-Slater (HFS) method.  The bound state and continuum wavefunctions from the HFS method are used to compute the cross sections of photoionization, shakeoff, electron impact ionization and electron-ion recombination processes and the decay rates of Auger and fluorescence decay processes in all participating ECs. The molecular dynamics method tracks the dynamics of ions and delocalized ionized electrons.

A classical over-the-barrier model simulates electron transfer dynamics in the valence shell.  In this model, electron transfer is allowed to instantaneously fill a vacancy in the valence shell of a neighboring atom when the electron binding energy is higher than the Coulomb barrier.  When the atoms are far apart, the resulting Coulomb barrier will suppress electron transfer.  The sensitivity of the charge transfer dynamics with respect to the bond distance has been observed experimentally in x-ray excited iodomethane \cite{Erk2014:science}. Our combined MC/MD and COB method allows tracking all decay, electron redistribution and fragmentation pathways for all breakup modes simultaneously using one calculation.  At the same time, the calculation computes ion charge states and dissociation energies for comparison with the experimental measurements. Our model does not include interatomic Coulombic decay and electron-transfer-mediated decay processes that are discussed in Ref.~\cite{Jahnke2020:cr}.

In this paper, we calculate the relaxation and fragmentation pathways of a single core-hole vacancy in IBr.  The vacancy is in the 2p subshell of either Br or I.  For both calculations, we used 10$^{6}$ trajectories to capture the complex decay landscape and fragmentation dynamics.

\section{RESULTS AND DISCUSSION\label{res}}

\subsection{Br 1s excitation}


\begin{figure}
\includegraphics[width=6.5in]{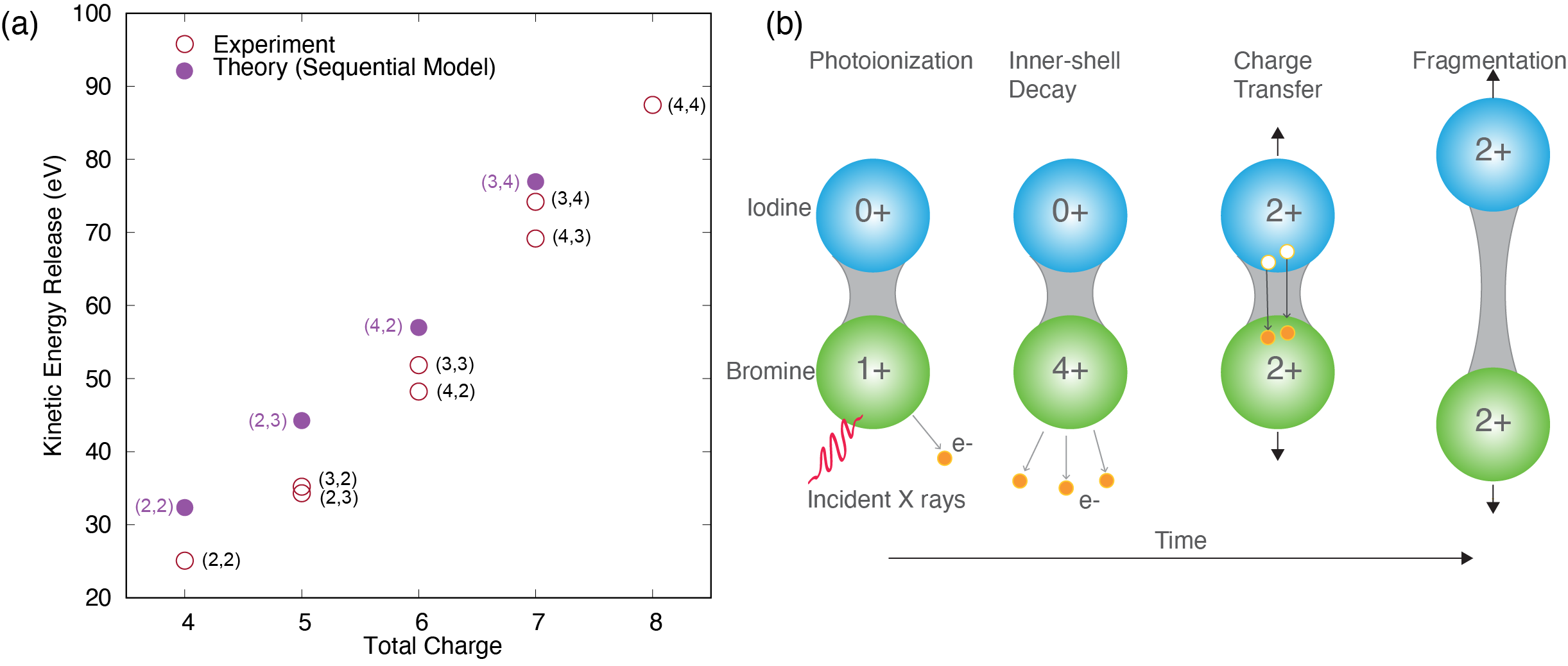}
\caption{(Color online) (a) Kinetic energy distributions of various ion fragmentation pairs of IBr measured at 13486 eV in coincidence with Br $K\alpha$ emission.  The open and filled circles are the experimental data and calculated results from a sequential model, respectively. (b) A schematic of the sequential model of inner-shell deexcitation and fragmentation dynamics of IBr resulting in the Br$^{2+}$/I$^{2+}$ ion.}
\label{IBr_BrKa_sequential}
\end{figure}


Figure \ref{IBr_BrKa_sequential}(a) shows the kinetic energy releases of IBr as a function of the total charge, where Br 1s is ionized followed by $K\alpha$ x-ray detection in coincidence. We note that our experimental results include both isotopes of Br.  Figure \ref{IBr_BrKa_sequential} shows the data for $^{79}$Br.  The data for $^{81}$Br are in Table~\ref{table2}.   To understand these data, we used two models.  The first model is called the sequential model and an example is depicted in Fig. \ref{IBr_BrKa_sequential}(b). In this model, electron transfer processes can only take place when the Br atom is fully relaxed and all the vacancies in Br are in the valence shell. Then, a Coulomb explosion fragments the molecule.   Figure \ref{IBr_BrKa_sequential}(a) plots the calculated average kinetic energy of various fragmentation channels, indicated by ($q_{Br}$,$q_{I}$), where $q_{Br}$ and $q_{I}$ are the final charge states of Br and I ions, respectively.  The agreement between experiment and this model is not good.   In particular, the sequential model suppresses (3,2), (3,3), (4,3) and (4,4) breakup modes.  To understand the reason behind the suppression, let us consider the two breakup modes, (3,2) and (2,3), for the total charge state of 5+.  Both modes start from the molecular electronic configuration (EC) of Br$^{5+}$/I in the initial molecular geometry and they are produced after 4 Auger decays in Br. Then, two electron transfer processes take place to generate the Br$^{3+}$/I$^{2+}$ molecular ion.   Since the binding energy of the iodine valence electron in Br$^{3+}$/I$^{2+}$ is still higher than the Coulomb barrier, an additional electron transfer process can take place to convert Br$^{3+}$/I$^{2+}$ to Br$^{2+}$/I$^{3+}$. As a result, the (3,2) mode is suppressed.

\begin{figure}
\includegraphics[width=6.5in]{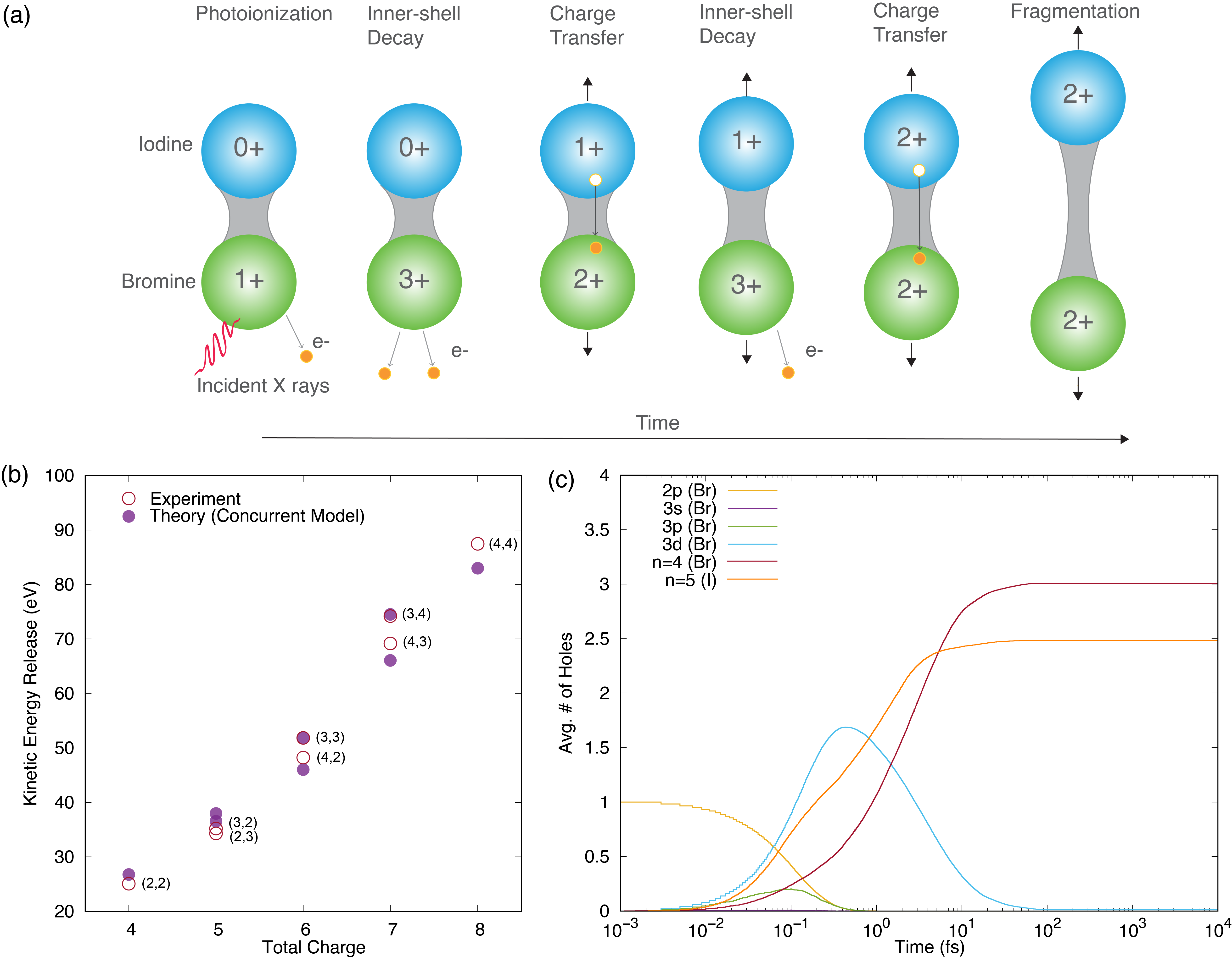}
\caption{(Color online) (a) Concurrent model of inner-shell deexcitation and fragmentation dynamics of IBr. (b) Kinetic energy distributions of various ion fragmentation pairs of IBr measured at 13486 eV in coincidence with Br $K\alpha$ emission.  The open and filled circles are the experimental data and calculated results from a concurrent model, respectively. (c) Average number of vacancies in each subshell of Br and the valence shell of I as a function of time, starting from a vacancy in 2p of Br.}
\label{ConcurrentModel}
\end{figure}



We then consider a second model in which electron transfer, Auger decay and fragmentation dynamics can take place concurrently.  Here, we can have the situation where Br is partially relaxed with vacancies in 3d and valence shells, and electron transfer can take place, as shown in Fig. \ref{ConcurrentModel}(a).  Our calculated results from the concurrent model agree well with the experimental data, as shown in Fig. \ref{ConcurrentModel}(b).  


Figure \ref{ConcurrentModel}(c) is a plot of the average number of vacancies in various subshells of Br and the valence shell (n=5) of I as a function of time, starting from a Br 2p vacancy.  It shows that the decay of a 2p hole is followed by the creation and decay of multiple 3d holes.  Then, an electron transfer, indicated by the orange line, takes place.  Our results show that the cascade is a multistep process across multiple timescales (attoseconds to $>$10 picoseconds) to create multiply charged ions.  One important result is that the decay of 3d vacancies and electron transfer both take place at the femtosecond timescale, supporting the concurrent model. The fluorescence steps that transfer the electrons from 4s to 4p, which are not shown in Fig. \ref{ConcurrentModel}(c), take picoseconds and longer.


\begin{figure}
\includegraphics[width=6.in]{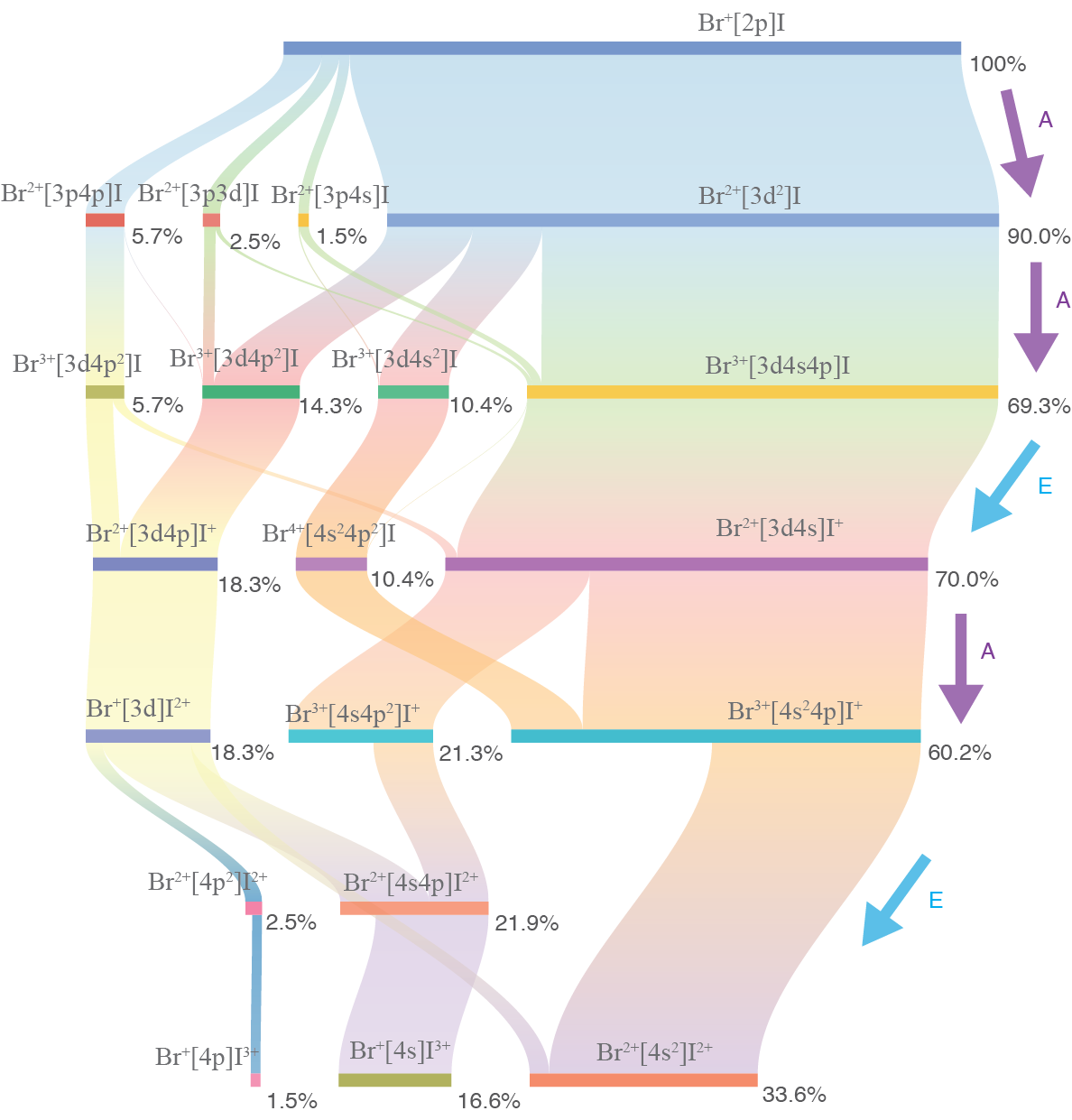}
\caption{(Color online) Sankey diagram showing the probability of the important electronic configurations participating in the decay of a core-excited IBr molecule leading to the breakup modes associated with a total charge of 4.  The horizontal bars indicate various electronic configurations, in which the bracket shows the vacancies in each Br subshell compared to its ground state of the Br atom, whereas the vacancies in the I are in its valence shell.  The width of each bar shows the cumulative probability, indicated by the percentages, during the decay. The set of arrows on the right side of the plot depicts the most dominant pathways leading to production of Br$^{2+}$/I$^{2+}$ and Br$^{3+}$/I$^{+}$, in which the purple and blue arrows represent Auger decay and electron transfer processes. The diagram does not plot the last fluorescence steps that fill the 4s vacancies in the cascades.}
\label{Sankey}
\end{figure}

\begin{figure}
\includegraphics[width=3.5in]{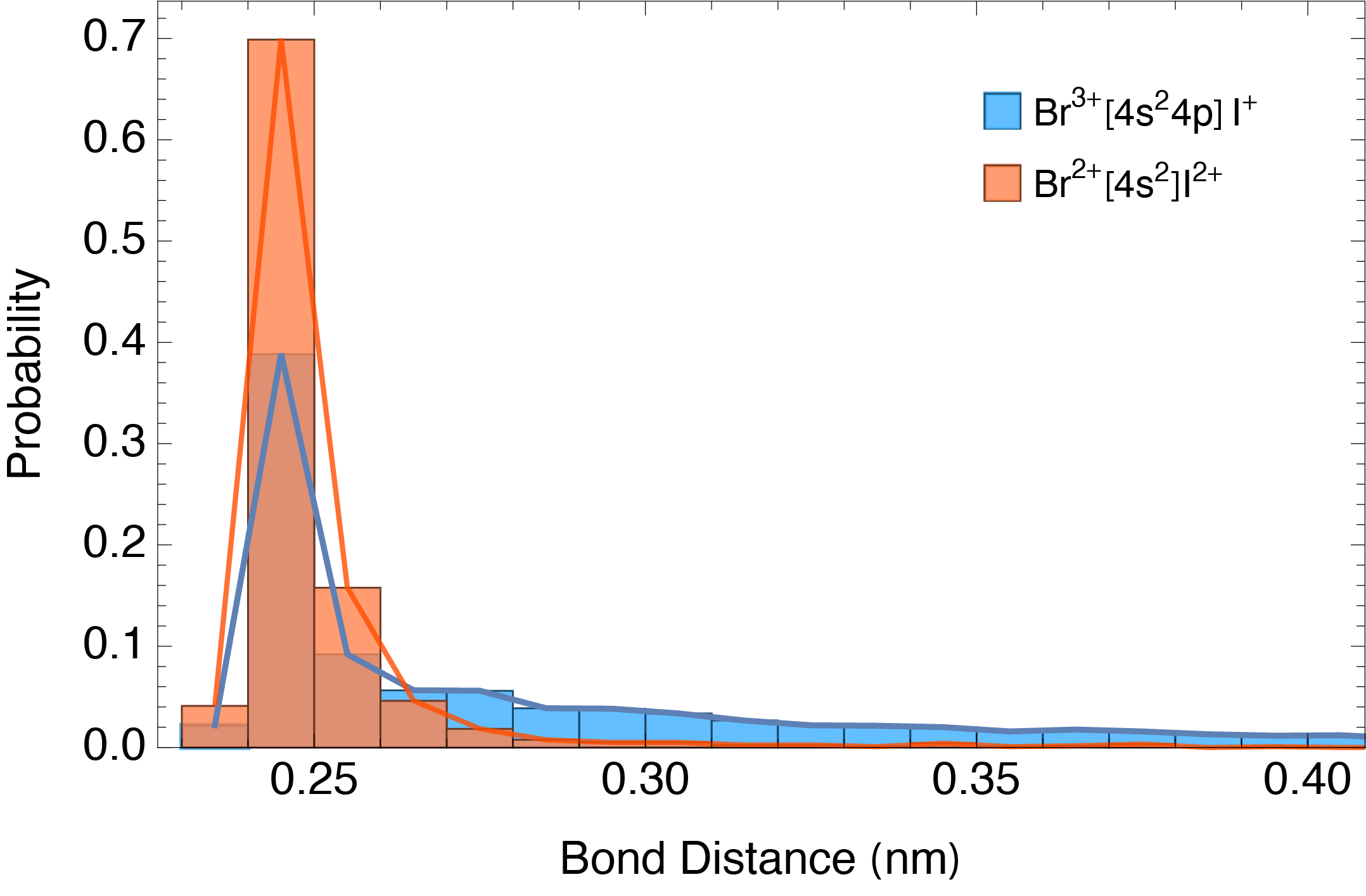}
\caption{(Color online) Distribution of IBr bond distance for two electronic configurations, Br$^{3+}$[4s$^2$4p]I$^{+}$ and Br$^{2+}$[4s$^2$]I$^{2+}$, when they are produced. The blue and orange solid lines trace out the distribution for Br$^{3+}$[4s$^2$4p]I$^{+}$ and Br$^{2+}$[4s$^2$]I$^{2+}$, respectively.}
\label{BondDistance}
\end{figure}

The inner-shell decay is rather complex, as shown in the Sankey diagram in Fig. \ref{Sankey}.  This diagram plots the important electronic configurations (ECs) for the three breakup modes (3,1), (2,2) and (1,3) with a total charge of 4+.  The individual bars indicate different ECs and the widths show their cumulative probabilities.  We highlight one dominant pathway that leads to (3,1) and (2,2) breakup modes from Fig. \ref{Sankey}.  Starting from a configuration with a Br 2p vacancy in IBr, 90\% of the 2p hole configuration will decay into a EC (Br$^{2+}$[3d$^2$]I) with two 3d holes.  Then, the decay of a 3d hole creates two additional holes in the valence shells of Br.  After that an electron transfer moves an electron from I to Br and creates Br$^{2+}$[3d4s]I$^{+}$ that can initiate charge separation/bond elongation dynamics due to Coulombic repulsion.  This electron transfer is then followed by the decay of the remaining 3d hole to produce a EC of Br$^{3+}$[4s$^2$4p]I$^{+}$.  We find that the timing of the decay of this 3d hole with respect to the bond elongation process can lead to two different breakup modes.  Figure \ref{BondDistance} shows that Br$^{2+}$[4s$^2$]I$^{2+}$ is produced over a narrow range of bond distances that are close to the initial bond distance of 0.2469 nm, while Br$^{3+}$[4s$^2$4p]I$^{+}$ is produced over a broader range of bond distances. Figure \ref{BondDistance} also reveals that the probability of producing Br$^{2+}$[4s$^2$]I$^{2+}$ drops rapidly with increasing bond distance.  This suggests that only molecules (with EC of Br$^{3+}$[4s$^2$4p]I$^{+}$) that have decay of the 3d hole taking place before sufficient bond extension can undergo an additional electron transfer to produce Br$^{2+}$[4s$^2$]I$^{2+}$, which leads to the (2,2) mode.  On the other hand, electron transfer is suppressed in molecules that undergo the decay after a substantial bond extension.  As a result, these molecules remain in the EC of Br$^{3+}$[4s$^2$4p]I$^{+}$ and lead to the (3,1) mode.  The depicted Sankey diagram does not show the last fluorescence steps that fill the remaining one or two 4s vacancies during the cascade.  Using A, F and E to symbolize Auger, fluorescence, electron transfer processes, the typical sequences for (2,2) and (3,1) are AAEAEFF and AAEAFF, respectively.

Interestingly, Fig. \ref{BondDistance} shows that $\sim$5\% of the Br$^{3+}$[4s$^2$4p]I$^{+}$ and Br$^{2+}$[4s$^2$]I$^{2+}$ populations has bond length shorter than the initial bond length (0.2469 nm) of the molecules.  Our analysis suggests that at least two factors can initiate nuclear motion that leads to bond contraction, including ion recoil during the emission of electrons and Coulomb attraction between ions and continuum electrons \cite{Ferguson2016:SA}. Future calculations with accurate treatment of the molecular effects for electron transfer processes and fragmentation dynamics will be insightful to validate the prediction of the bond contraction.

\subsection{I 1s excitation}
\begin{figure}
\includegraphics[width=6.5 in]{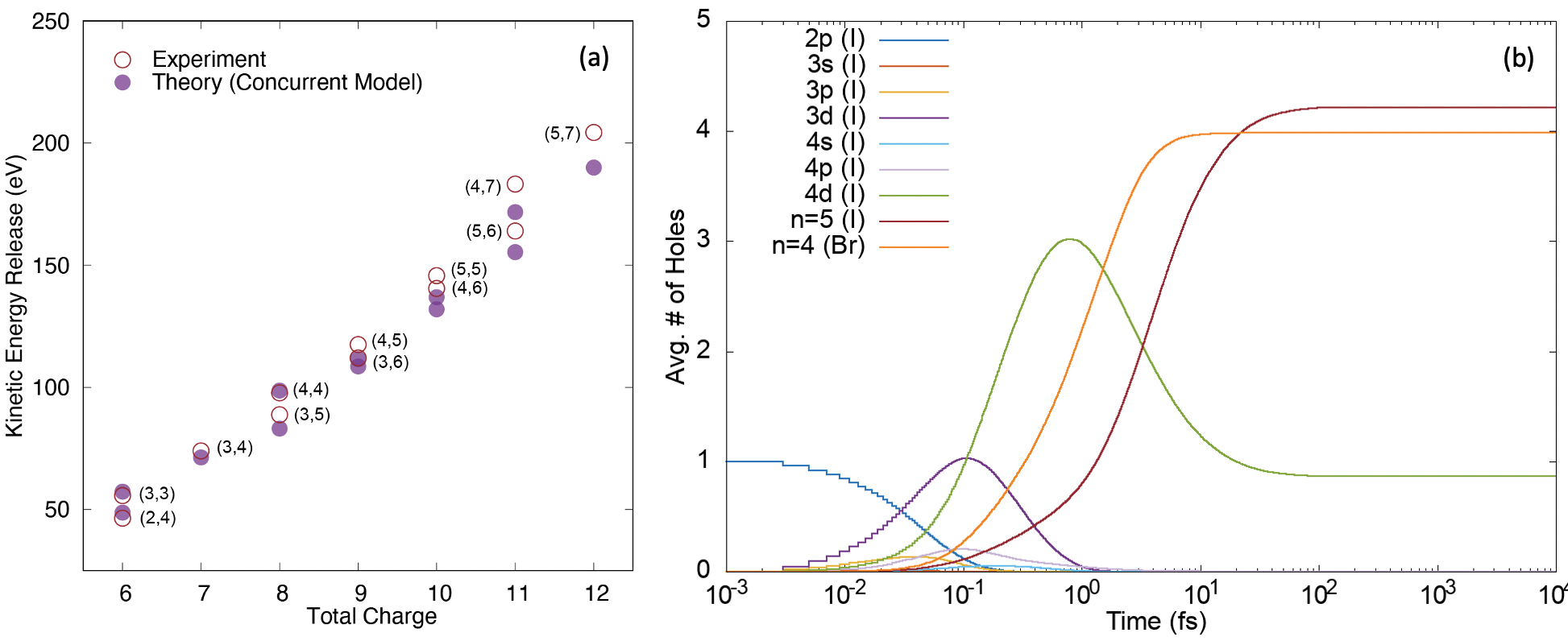}
\caption{(Color online) (a) Kinetic energy distribution of various ion fragmentation pairs of IBr measured at 33194 eV in coincidence with I $K\alpha$ emission.  The open and filled circles are the experimental data and calculated results from a concurrent model, respectively. (b) Average number of vacancies in each subshell of I and the valence shell of Br as a function of time, starting from a vacancy in 2p of I.}
\label{IBr_IKa}
\end{figure}

In addition to the data from Br excitation, we have also collected data associated with I excitation in coincidence with I $K\alpha$ emission, as shown in Fig. \ref{IBr_IKa}.  We found that the calculated results from the concurrent model agree with the experimental data.  Our calculation shows that the electron transfer processes and the decay of I 4d vacancies both take place at the femtosecond timescale, supporting the concurrent model.  
In our experimental data, two break-up modes (4,5) and (3,6) are identified for ion pairs with a total charge state of 9+.  Our calculation can illuminate the dominant mechanisms associated with these two break-up modes.  

A typical sequence leading to the (4,5) mode starts with the iodine atom being rapidly charged to either 5+ or 6+ via multiple Auger decays, which create multiple vacancies in 4d and valence shells (n=5). This is followed by 4 electron transfers to produce charged bromine and iodine ions, which will initiate molecular fragmentation due to Coulomb repulsion.  In some sequences, an Auger decay can take place between the electron transfer events.  As the ions separate, the iodine atom will undergo 2 or more Auger decays to further charge up the iodine ion.  These Auger events are sometimes followed by fluorescence events, which can take place at times longer than 10 ps after the start of the cascade to create a 5+ iodine ion with all vacancies in the 5p subshells. Using A, F and E to represent Auger, fluorescence and electron transfer processes, some example sequences for the (4,5) break-up mode are AAAAAEAEEEAAF and AAAAAAEEEEAAF.  For the (3,6) breakup mode, typical example sequences include AAAAAEAEEAAF and AAAAAAEEEAAF, in which the cascades are similar to those found in the (4,5) modes except the (3,6) mode involves only 3 electron transfers.  




\section{CONCLUSION\label{con}}

We presented a combined experimental and theoretical study of inner-shell cascade dynamics of IBr molecules following 1s ionization at the I and Br $K$-edges.  Using x-ray/ion coincidence spectroscopy, we identified the charge states and kinetic energies of two correlated Br and I fragment ions associated with the cascades from a 2p vacancy in I and Br.  

Our MC/MD+COB calculation suggests that a model that allows inner-shell decays, electron transfer and dissociation to occur concurrently is necessary to account for all experimentally observed breakup modes and to show good agreement with the measured ion kinetic energies.  Our calculations enable tracking of the multistep cascades, depicting the charging process and identifying the dominant pathways and participating transient electronic states spanning across multiple time scales from attoseconds to picoseconds. Due to the finite lifetimes of the core-excited transient states, the resulting charging process does not occur instantaneously.  Our analysis further shows that the timing of the Br 3d and I 4d hole decays during the fragmentation can affect the probability of the subsequent electron transfer processes and the production of particular breakup modes.

Future x-ray/ion, x-ray/electron, and electron/ion coincidence spectroscopy studies that selectively probe the decay dynamics starting from specific initial states in heavy-element containing molecules will be insightful.  On the theory side, our current model provides an efficient way to simulate the multistep and multielectron cascade process based on a probabilistic electron transition and localized electron and hole description.  Our model can be improved to explicitly compute the probability of electron transfer that goes beyond the instantaneous model and includes tunneling.  Future theoretical methods based on molecular wavefunctions will be important to access the effects of electronic coherence and molecular orbitals and symmetry, in particular the question of hole/electron localization and nuclear motion (such as bond contraction) during the vacancy cascade.


\section{ACKNOWLEDGEMENTS}
This work was supported by the U.S. Department of Energy, Office of Science, Basic Energy Sciences, Chemical Sciences, Geosciences, and Biosciences Division. Use of the Advanced Photon Source, an Office of Science User Facility operated for the U.S. Department of Energy (DOE) Office of Science by Argonne National Laboratory, was supported by the U.S. DOE under Contract No. DE-AC02-06CH11357.

\end{document}